\documentclass[aps,twocolumn,a4paper,amsmath,showpacs,floatfix]{revtex4}
\usepackage{graphicx}
\usepackage{amssymb}
\usepackage{amscd}
\begin{document}

\title{Demixing behavior in two-dimensional mixtures of anisotropic hard bodies} 
\author{Yuri Mart\'{\i}nez-Rat\'on}
\email{yuri@math.uc3m.es}

\affiliation{Grupo Interdisciplinar de Sistemas Complejos (GISC),
Departamento de Matem\'aticas, Escuela Polit\'ecnica Superior,
Universidad Carlos III de Madrid,
Avenida de la Universidad 30, E-28911 Legan\'es, Madrid, Spain.
}

\author{Enrique Velasco}
\email{enrique.velasco@uam.es}

\affiliation{Departamento de F\'{\i}sica Te\'orica de la Materia Condensada
and Instituto de Ciencia de Materiales Nicol\'as Cabrera,
Universidad Aut\'onoma de Madrid, E-28049 Madrid, Spain.}

\author{Luis Mederos}
\email{l.mederos@icmm.csic.es}

\affiliation{Instituto de Ciencia de Materiales, Consejo Superior de
Investigaciones Cient\'{\i}ficas, E-28049 Cantoblanco, Madrid, Spain.}

\date{\today}

\begin{abstract}
Scaled particle theory  for a binary mixture of 
hard discorectangles and for a binary mixture of hard rectangles
is used to predict possible liquid-crystal 
demixing scenarios in two dimensions. Through a 
bifurcation analysis from the isotropic phase, it is shown 
that isotropic-nematic demixing is possible in two-dimensional liquid-crystal mixtures 
composed of hard convex bodies. This bifurcation analysis is 
tested against exact calculations of the phase diagrams in the framework 
of the restricted-orientation two-dimensional model (Zwanzig model). Phase diagrams 
of a binary mixture of hard discorectangles are calculated through the 
parametrization of the orientational distribution functions. The results 
show not only isotropic-nematic, but also nematic-nematic demixing
ending in a critical point, as well as an isotropic-nematic-nematic 
triple point for a mixture of hard disks and hard discorectangles.

\end{abstract}

% 64.70.Md  Transitions in liquid crystals
% 64.75.+g  Solubility, segregation, and mixing; phase separation
% 61.20.Gy  Theory and models of liquid structure

\maketitle

\section{Introduction} 
The demixing behavior of hard-core three-dimensional 
additive mixtures composed of 
particles without orientational degrees of freedom, in particular 
the hard-sphere and parallel hard-cube systems, has been recently studied 
in depth by several authors using Monte Carlo 
simulation techniques \cite{Dijkstra1,Almarza,Buhot} and
theory \cite{Shuri,Mederos}. The main conclusions 
that can be drawn from these studies is that fluid-fluid demixing 
is always metastable with respect to fluid-solid demixing for 
asymmetric mixtures in which the solid phase is composed of big particles 
and the fluid phase is enriched in small particles 
(demixing behavior similar to the one found in mixtures 
of parallel hard cubes 
on a cubic lattice \cite{Lafuente}).
The physics behind this 
demixing behavior is known as the depletion effect \cite{Biben}, and can 
be explained as the effective attraction between two large particles due to 
the uncompensated osmotic pressure exerted by the 
small particles when the two excluded volumes between 
big and small particles overlap. Thus in the demixed phases enriched by each of the 
species the accessible volumes to the particles is maximized and as a consequence the 
total configurational entropy is increased.

Three-dimensional mixtures of additive anisotropic particles 
possess a demixing behavior 
which differs from that of mixtures composed of isotropic  
particles because of the presence of particle orientational 
degrees of freedom. It has been shown 
that demixing strongly depends on the shape of the 
particles (spherical, oblate or prolate) \cite{Roij1,Wensink,Perera,Varga1,
Martinez-Raton1,Schmidt,Oversteegen} 
and for a particular geometry, on its aspect ratio (the ratio 
between the characteristic lengths of particles)  
\cite{Roij2,Dijkstra2,Varga2}. 
For example, one of the
demixed phases can be an oriented phase, i.e., a nematic (N) phase
where the particles are aligned on average along the nematic director. 
It is easy to predict the most common demixing scenario, it will proceed
between a N phase composed of the longer 
particles and an isotropic (I) phase composed of the short-particle component. 
In this case the excluded volume 
between the large particles is minimized by their parallel alignment.  
I-I and N-N 
demixing was also found in mixtures of anisotropic 
particles \cite{Roij1,Wensink,Perera,Varga1,Martinez-Raton1,
Schmidt,Oversteegen,Roij2,Dijkstra2,Varga2}, showing that fluid-fluid 
demixing is a common scenario in these mixtures, which has recently
been confirmed experimentally \cite{Purdy}.

The scaled particle theory (SPT) was initially developed for hard 
spheres \cite{Reiss} and later extended to anisotropic particles 
\cite{Cotter,Lasher,Barboy}.
The usual formulation of SPT
for mixtures of 
hard convex bodies restricted to the isotropic orientational 
phase 
has as a main ingredient the expression for the second 
virial coefficients as a function of the volumes, surface 
areas, and mean curvatures of hard convex bodies \cite{Isihara, 
Kihara}. This  
exact result obtained in the 1950's \cite{Isihara} was 
used to show that in two dimensions
I-I demixing is not allowed \cite{Talbot}, the gain in accessible 
volume after demixing is much lower in two dimensions, which explains the 
stability of mixtures with respect to I-I demixing 
at any composition. 
This result has been 
confirmed by other theories constructed from the expression of the 
direct correlation function in terms of the geometric measures of the
particles \cite{Perera}, in the same spirit as the fundamental measure theory (FMT) 
for hard spheres \cite{Rosenfeld,Tarazona}. 
Some authors have studied the possibility of 
demixing in a mixture of perfectly oriented two-dimensional additive particles 
\cite{Perera}, arriving at the same conclusion, the mixture is always stable.  
However, in contrast to these findings, we will show in section III that 
the SPT approximation does predict N-N demixing in the limit of parallel alignment.
Also, the previous negative results cannot be taken as definite, since
the suppression of orientational fluctuations is a crude approximation to study the 
possibility of  N-N demixing, so that the question about the existence of  
demixing (I-N or N-N) in two dimensions 
is still open. Note that we explicitly distinguish I-N demixing from 
the usual orientational I-N phase transition. The reason for this will 
be explained later in Sec. IV C.

As opposed to additive mixtures, nonadditive mixtures  
of hard disks can demix, as was shown by several authors \cite{Tenne,Mountain}.
This result is not surprising since the demixed phases, composed 
of practically the 
same kind of species, are approximately additive and, as a consequence, have 
less excluded volumes between any pair of particles compared to the mixed 
state. Nonadditive mixtures made up of anisotropic particles have not been 
analyzed so far. The coupling between nonadditiveness in the interactions and
orientational degrees of freedom is expected to give rise to interesting 
phenomenology, which is worth investigating. We will not pursue these aspects
in the present paper.

The purpose of this work is to shed light on the question 
about the existence of I-N or N-N demixing 
in two-dimensional mixtures of additive 
hard anisotropic particles. We have used the 
SPT formalism specified for a mixture of 
hard rectangles (HR) and hard discorectangles (HDR), and 
applied a bifurcation analysis from the isotropic phase to find the 
possibility of I-N demixing. This analysis has been shown to be useful in 
the study of phase transitions with symmetry breaking. In particular 
it was used to study the I-N transition in two dimensions in a
one-component system of hard needles  \cite{Raveche}, and in a system composed 
of self-assembled rods \cite{Schoot} in the same two-dimensional Onsager limit.
Using the bifurcation analysis, we have found that, even for a system of HDR 
(where the I-N transition is continuous in the one-component limit, as indicated 
by simulation \cite{Frenkel} and density-functional theory \cite{Martinez-Raton2}) 
the mixture can demix in two phases of different composition. In addition, we 
have calculated the phase diagrams of different mixtures of HDR 
and confirmed this demixing scenario. In some mixtures we have 
also found a N-N phase separation which ends in a critical point,
as well as a triple coexistence between an I phase and 
two different nematics.  

The paper is organized as follows. In Sec. II we present the  
theoretical model, and in Sec. III the bifurcation analysis. The 
results from this analysis are shown for HDR in Sec. 
IV A, for HR in Sec. IV B, and a check using a very simple 
model (Zwanzig approximation) is presented in Sec. IV C. 
Section V shows the results from the calculations 
of the phase diagrams of HDR mixtures. Finally some  
conclusions are drawn in Sec. VI.

\section{Model}

The key quantity in the development 
of the SPT for a mixture of hard convex bodies in two dimensions (see Ref.\cite{Schaklen})
is the averaged (over all possible orientations of a fluid particle of 
species $\nu$) 
excluded area between an inserted scaled particle $s$ with orientation 
$\phi_1$ (measured from the nematic director) and a fluid particle 
of species $\nu$ with orientation $\phi_2$, i.e., 
\begin{eqnarray}
\langle V_{\rm{excl},\nu}^{\alpha}\rangle(L_s,\sigma_s,\phi_1)
=\int d\phi_2 h_{\nu}(\phi_2)
V_{\rm{excl},\nu}^{\alpha}(L_s,\sigma_s,\phi_{12}),\hspace*{0.1cm} 
\end{eqnarray}
where $L_s$ and $\sigma_s$ are the length and width of the scaled particle, 
$h_{\nu}(\phi_2)$ is the orientational distribution function of species 
$\nu$ and $\phi_{12}=\phi_1-\phi_2$ the relative angle between the 
axes of particles $s$ and $\nu$. The superindex $\alpha$ labels 
the nature of the particle, either hard rectangles ($\alpha=\text{HR}$) or 
hard discorectangles ($\alpha=\text{HDR}$). The 
excluded area between two rectangles ($s$ and $\nu$) is 
\begin{eqnarray}
V^{\rm{HR}}_{\rm{excl},\nu}(L_s,\sigma_s,\phi_{12})&=&
\left(L_{\nu}L_s+\sigma_{\nu}\sigma_s\right)
|\sin \phi_{12}|+v_{\nu}+v_s\nonumber \\ &+& 
\left(L_{\nu}\sigma_s+L_s\sigma_{\nu}\right)|\cos \phi_{12}|.
\end{eqnarray} 
For hard discorectangles
\begin{eqnarray}
V^{\rm{HDR}}_{\rm{excl},\nu}(L_s,\sigma_s,\phi_{12})&=&
L_{\nu}L_s|\sin\phi_{12}|+v_{\nu}+v_s\nonumber\\
&+&L_{\nu}\sigma_s+L_s\sigma_{\nu}+\frac{\pi}{2}\sigma_{\nu}\sigma_s,
\end{eqnarray}
where for HR $v_{\beta}=L_{\beta}\sigma_{\beta}$ is the area of 
species $\beta$ ($\beta=\{\nu,s\}$) while for HDR $v_{\beta}=
L_{\beta}\sigma_{\beta}+\pi\sigma_{\beta}^2/4$. 
The reversible work required to insert the scaled particle 
with fixed orientation 
coincides with the excess chemical potential and, in the limit of small sizes 
($L_s\ll L_{\nu}$, $\sigma_{s}\ll \sigma_{\nu}$ for any $\nu$), it has the 
following asymptotic form \cite{Reiss}:
\begin{eqnarray}
\beta \mu_{\rm{exc}}(\phi_1)&\sim& \mu^{(0)}(L_s,\sigma_s,\phi_1)
\nonumber \\&\equiv&-\ln\left[1-\sum_{\nu}\rho_{\nu}\langle V_{\rm{excl},\nu}^{\alpha}\rangle 
(L_s,\sigma_s,\phi_1)\right],\hspace*{0.2cm}
\end{eqnarray}
where $\rho_{\nu}$ is the density of species $\nu$ and the sum runs over 
all species. In the opposite limit of large sizes ($L_s\gg L_{\nu}$, 
$\sigma_{s}\gg\sigma_{\nu}$) this work coincides with the thermodynamic 
work required to open a cavity of area $v_s$, which is equal to $P v_s$, 
where $P$ is the fluid pressure. The SPT interpolates between both limits 
using a Taylor expansion of the function $\mu^{(0)}(L_s,\sigma_s,\phi_1)$ 
around the value $(L_s,\sigma_s)=(0,0)$. The second term of this expansion 
is fixed to $Pv_s$. Finally, all the particle lengths are taken to be those 
of any one of the species, say $\nu$, which results in 
\begin{eqnarray}
&&\beta \mu_{\rm{exc},\nu}(\phi_1)=-\ln(1-\eta)\nonumber \\
&+&\frac{\sum_{\tau} \rho_{\tau}\int d\phi_2 h_{\tau}(\phi_2)
V^{(0)}_{\nu\tau}(\phi_{12})}{1-\eta}
+\beta P v_{\nu},
\end{eqnarray}
where $\eta=\sum \rho_{\nu}v_{\nu}$ is the total packing fraction and 
$V^{(0)}_{\nu\tau}(\phi_{12})=V_{\rm{excl},\nu}^{\alpha}(L_{\tau},\sigma_{\tau},
\phi_{12})-v_{\nu}-v_{\tau}$. The excess chemical potential of species 
$\nu$ is the angular average 
\begin{eqnarray}
\beta \mu_{\rm{exc},\nu}&=&\int d\phi_1 h_{\nu}(\phi_1) \left[
\beta \mu_{\rm{exc},\nu}(\phi_1)\right]=-\ln(1-\eta)\nonumber \\&+&
\frac{\sum_{\tau}\rho_{\tau}\langle \langle V^{(0)}_{\nu\tau}
\rangle\rangle}
{1-\eta}+\beta P v_{\nu},
\label{pote}
\end{eqnarray}
where $\langle\langle\cdots\rangle\rangle$ means the following 
double angular average:
\begin{eqnarray}
\langle \langle V^{(0)}_{\nu\tau}\rangle\rangle=
\int d\phi_1 h_{\tau}(\phi_1)\int d\phi_2 h_{\nu}(\phi_2) 
V^{(0)}_{\nu\tau}(\phi_{12}).
\end{eqnarray}

Integrating the thermodynamic relations 
\begin{eqnarray}
\frac{\partial \beta P}{\partial\rho_{\nu}}=1+
\sum_{\tau}\rho_{\tau}\frac{\partial\beta\mu_{\rm{exc},\tau}}
{\partial \rho_{\nu}},\nonumber
\end{eqnarray}
with the use of Eq. (\ref{pote}) allows us to find 

\begin{eqnarray}
\beta P=\frac{\rho}{1-\eta}+\frac{1}{2}\frac{\sum_{\nu\tau}
\rho_{\nu}\rho_{\tau}\langle\langle V_{\nu\tau}^{(0)}\rangle\rangle}
{(1-\eta)^2},
\label{presion}
\end{eqnarray}
where $\rho=\sum_{\nu}\rho_{\nu}$ is the total density.
Finally, through the definition of the pressure 
$\beta P=\rho+\sum_{\nu}\rho_{\nu}\left[\beta \mu_{\rm{exc},\nu}\right]-
\Phi_{\rm{exc}}$, where $\Phi_{\rm{exc}}=\beta {\cal F}_{\rm{exc}}/V$ is 
the excess part of the free energy density in reduced units, the result 
(\ref{presion}), and  
Eq. (\ref{pote}) we obtain 

\begin{eqnarray}
\Phi_{\rm{exc}}=-\rho \ln(1-\eta)+\frac{1}{2}\frac{\sum_{\nu\tau}
\rho_{\nu}\rho_{\tau}\langle\langle V_{\nu\tau}^{(0)}\rangle\rangle}
{1-\eta}.
\end{eqnarray}

The ideal part of the free energy density of the mixture is 
\begin{eqnarray}
\Phi_{\rm{id}}=\sum_{\nu} \rho_{\nu}\left(\ln \rho_{\nu} -1+
\int_0^{\pi} d\phi h_{\nu}(\phi)\ln\left[ \pi h_{\nu}(\phi)\right]\right),
\hspace*{0.1cm}
\end{eqnarray}
where all distribution functions are normalized as $\int_0^{\pi}d\phi
h_{\nu}(\phi)=1$ (note that, in view of the head-tail symmetry of 
the particles, the angle $\phi$ can be restricted to the interval 
$[0,\pi]$). The functional minimization of 
$\Phi=\Phi_{\rm{id}}+\Phi_{\rm{exc}}$ 
with respect to the $\{h_{\nu}\}$ allows, as usual, to find
the equilibrium distribution 
functions and correspondingly the equilibrium free energy of the mixture.

The above theoretical scheme will be used in the following section to develop 
a bifurcation analysis of I-N demixing for HDR and HR. Also it will be used 
to calculate the phase diagram of mixtures. In order to do that 
we need to fix the fluid pressure, 
which means that the composition of one of the 
species and the total fluid density are no longer independent variables. Once
the independent variable is chosen, the dependent variable can be calculated through  
the constant pressure criterion. Thus, the 
adequate thermodynamic potential to work with is the Gibbs free energy 
per particle in reduced units
$\beta g=\left(\Phi+\beta P\right)/\rho$. The minimization of $\Phi$ with 
respect to the order parameters (in the case of the nematic phase), 
the condition of constant pressure and the double tangent construction 
on $\beta g$ with respect to the composition of one of the species, 
allows us to calculate the coexistence condition between different phases. 
Changing the pressure and repeating the above steps we have found the 
phase diagrams of different binary mixtures.

\section{Bifurcation analysis}

The usual bifurcation analysis for a one-component fluid with symmetry 
breaking includes (i) an order parameter expansion of the free energy around 
the bifurcation point; 
(ii) the calculation of the inverse isothermal 
compressibility of the ordered phase at the same point. The order of the 
phase transition can be elucidated by the combined use of both criteria, 
the sign of the free-energy difference between the 
ordered and disordered phases,
already minimized with respect to the order parameter, 
and the sign of the isothermal compressibility of the ordered phase,  
evaluated at bifurcation. If the system exhibits a tricritical point, its  
location can be obtained from the vanishing of either the first coefficient in  
the expansion of the free-energy difference 
or the isothermal compressibility,  depending on which of them occurs first 
\cite{Sluckin}.

For binary mixtures fluid-fluid demixing without symmetry breaking is usually 
calculated from the vanishing of the determinant of the matrix 
with elements $\partial^2\Phi/\partial\rho_i\partial\rho_j$ which means that 
the stability of the mixture with respect to volume and composition 
fluctuations is violated. This allows us to obtain the demixing spinodal. 
However, in a symmetry-breaking transition the above matrix 
should be calculated  from the minimized free-energy  
of the ordered phase using 
the order-parameter expansion up to the order required. This criterion is 
equivalent to the loss of convexity of the Gibbs free energy per particle  
of the ordered phase 
with respect to the mixture composition at the bifurcation point.
Since our aim is the study of I-N demixing, which involves an orientational 
symmetry breaking, we will implement the latter scheme. 

A Fourier-series decomposition of the orientational distribution functions  
\begin{eqnarray}
h_{\nu}(\phi)=\frac{1}{\pi}
\left(1+\sum_{k\geq 1}h^{(\nu)}_k \cos (2k\phi)\right)
\end{eqnarray}
should retain 
only even harmonics, due to the symmetry of the particles studied here 
[$h_{\nu}(\phi)=h_{\nu}(\pi-\phi)$]. In the neighborhood 
of the I-N bifurcation point we can assume that the Fourier 
amplitudes $\{h^{(\nu)}_k\}$ are small.  
A Taylor expansion of the difference in
free energy per particle $\Delta \varphi=\varphi_{\rm{N}}-
\varphi_{\rm{I}}$ ($\varphi=\Phi/\rho$) 
between N and I phases, 
up to fourth order, is therefore valid (the order is 
defined by a small bifurcation parameter $\epsilon$, $h_k^{(\nu)}
\sim \epsilon^k$).  Up to fourth order, the ideal contribution to 
$\Delta\varphi$ reads 
\begin{eqnarray}
\Delta\varphi_{\rm{id}}&\approx& \sum_{\mu}x_{\mu}\Bigg(
\frac{1}{4}\left[\left(h_1^{(\mu)}\right)^2+\left(h_2^{(\mu)}\right)^2
\right]\nonumber\\&-&
\frac{1}{8}\left(h_1^{(\mu)}\right)^2h_2^{(\mu)}+
\frac{1}{32}\left(h_1^{(\mu)}\right)^4\Bigg),
\end{eqnarray}
whereas the excess contribution is 
\begin{eqnarray}
\Delta\varphi_{\rm{ex}}=\frac{1}{2}y\sum_{\mu\nu}x_{\mu}x_{\nu}
\frac{\Delta\langle\langle V_{\mu\nu}^{(0)}\rangle\rangle}{\langle v\rangle}.
\end{eqnarray}
We have defined 
\begin{eqnarray}
\Delta\langle\langle V_{\mu\nu}^{(0)}\rangle\rangle\equiv 
\langle\langle V_{\mu\nu}^{(0)}\rangle\rangle_{\rm{N}}-
\langle\langle V_{\mu\nu}^{(0)}\rangle\rangle_{\rm{I}},
\end{eqnarray}
with 
\begin{eqnarray}
\frac{\Delta\langle\langle V_{\mu\nu}^{(0)}\rangle\rangle}{\langle v\rangle}
=-\frac{1}{\pi}\sum_{k\geq 1}\frac{T_{\mu\nu}^{(k)}}{4k^2-1}h_k^{(\mu)}
h_k^{(\nu)},
\end{eqnarray}
where it is understood that the sum is to be truncated at fourth order 
in the expansion parameter. In the above expresions we have also defined
\begin{eqnarray} 
T_{\mu\nu}^{(k)}=\left\{\begin{array}{l}
\displaystyle{
\frac{L_{\mu}L_{\nu}}{\langle v\rangle}}, \hspace*{4.1cm} \text{for HDR},\\
\displaystyle{
\frac{\left[L_{\mu}+(-1)^k\sigma_{\mu}\right]
\left[L_{\nu}+(-1)^k\sigma_{\nu}\right]}{\langle v\rangle}}, \hspace*{0.1cm}
\text{for HR}\end{array}\right.
\end{eqnarray}
with $x_{\mu}=\rho_{\mu}/\rho$ being the molar fraction of species $\mu$, 
$y=\eta/(1-\eta)$, and $\langle v\rangle=\sum_{\nu}x_{\nu}v_{\nu}$ the 
average particle area of the mixture. The complete free-energy expansion 
up to fourth order reads

\begin{eqnarray}
\Delta \varphi&=&a_{11}^{(1)}\left(h_1^{(1)}\right)^2+a_{22}^{(1)}
\left(h_1^{(2)}\right)^2+2a_{12}^{(1)}h_1^{(1)}h_1^{(2)}\nonumber \\
&+&a_{11}^{(2)}\left(h_2^{(1)}\right)^2+a_{22}^{(2)}\left(h_2^{(2)}\right)^2
+2a_{12}^{(2)}h_2^{(1)}h_2^{(2)}\nonumber\\
&+&a_{11}^{(3)}\left(h_1^{(1)}\right)^2h_2^{(1)}+a_{22}^{(3)}
\left(h_1^{(2)}\right)^2h_2^{(2)}+a_{11}^{(4)}\left(h_1^{(1)}\right)^4\nonumber 
\\&+& a_{22}^{(4)}\left(h_1^{(2)}\right)^4.
\label{primero}
\end{eqnarray} 
The expressions for the $a_{\mu\nu}^{(k)}$'s are 
\begin{eqnarray}
a_{\mu\nu}^{(k)}&=&\frac{x_{\mu}}{4}\left(\delta_{\mu\nu}-\frac{2y x_{\nu}}
{(4k^2-1)\pi}T_{\mu\nu}^{(k)}\right), \label{first}\hspace*{0.1cm} k=1,2,\\
a_{\mu\mu}^{(3)}&=&-\frac{x_{\mu}}{8},\quad a_{\mu\mu}^{(4)}=\frac{x_{\mu}}
{32},\hspace*{1.4cm} \mu,\nu=1,2, \label{second} 
\end{eqnarray}
where $\delta_{\mu\nu}$ is the Kronecker delta. 

Minimizing $\Delta \varphi$ with respect to all the amplitudes except one  
(say $h_1^{(1)}$), substituting the results for the other 
amplitudes in (\ref{primero}), and 
neglecting all terms with order higher than four, we obtain an 
effective free energy difference as 
\begin{eqnarray}
\Delta \varphi=A \left(h_1^{(1)}\right)^2+B\left(h_1^{(1)}\right)^4,
\label{delta}
\end{eqnarray}
with the following explicit expressions for the coefficients $A$ and $B$:
\begin{eqnarray}
A&=&a_{11}^{(1)}-\frac{\left(a_{12}^{(1)}\right)^2}{a_{22}^{(1)}},\\
B&=&a_{11}^{(4)}+\left(a_{22}^{(4)}-\frac{\left(a_{22}^{(3)}\right)^2}
{4a_{22}^{(2)}}\right)\left(\frac{a_{12}^{(1)}}{a_{22}^{(1)}}\right)^4
\nonumber \\
&-&\frac{\displaystyle{\left[a_{11}^{(3)}-a_{22}^{(3)}\frac{a_{12}^{(2)}}
{a_{22}^{(2)}}\left(\frac{a_{12}^{(1)}}{a_{22}^{(1)}}\right)^2\right]^2}}
{\displaystyle{
4\left(a_{11}^{(2)}-\frac{\left(a_{12}^{(2)}\right)^2}{a_{22}^{(2)}}\right)}}.
\end{eqnarray}
Note that $A$ and $B$ depend on $x=x_2$ and $y$. To proceed we must 
bear in mind that demixing phase transitions usually
imply fractionation (different composition $x$ of the coexisting phases) 
as well as a change in packing fraction $\eta$. 
Therefore, we should expand both the composition $x$
and the variable $y$ around the bifurcation point $(x^*,y^*)$,
\begin{eqnarray} 
x\approx x^*+x^{(2)}\left(h_{1}^{(1)}\right)^2, \quad 
y\approx y^*+y^{(2)}\left(h_1^{(1)}\right)^2.
\label{expan}
\end{eqnarray}
Inserting the expansions (\ref{expan}) in the first derivative of 
(\ref{delta}) with respect to $h_1^{(1)}$ and equating the result to zero, 
order by order, we obtain the following conditions:  
\begin{eqnarray}
A^*&=&0,\quad {\cal O}\left[h_1^{(1)}\right],\label{uno}\\
A_x^*x^{(2)}+A^*_yy^{(2)}&=&-2B^*, \quad 
{\cal O}\left[\left(h_1^{(1)}\right)^3\right],
\label{dos}
\end{eqnarray}
where $f^*\equiv f(x^*,y^*)$ with ($f=A,B,A_x,A_y$), and the subindices 
$x,y$ in Eq. (\ref{dos}) mean the partial derivatives of $A$ with respect to $x$
and $y$, respectively. Solving Eq. (\ref{uno}) gives the packing fraction
as a function of composition at the bifurcation point, whereas
Eq. (\ref{dos}) allows to find a 
relation between $x^{(2)}$ and $y^{(2)}$, respectively.  

Expanding (\ref{delta}) up to fourth order around the bifurcation point 
and using (\ref{uno}) and (\ref{dos}) we obtain the 
energy difference as 
\begin{eqnarray}
\Delta \varphi =-B^* \left(h_1^{(1)}\right)^4. 
\label{difer}
\end{eqnarray}
Depending on the sign 
of $B^*$ the nematic branch bifurcates below (positive sign) or above 
(negative sign) the isotropic branch. The latter case corresponds to a 
first-order transition. Using (\ref{expan}) and (\ref{dos}), Eq. 
(\ref{difer}) can be rewritten as 
\begin{eqnarray}
\varphi_{\rm{N}}=\varphi_{\rm{I}}-\frac{1}{4B^*}\left[
A_x^*(x-x^*)+ A_y^*(y-y^*) \right]^2.
\label{uso}
\end{eqnarray}
The stability of the mixture with respect to volume and composition 
fluctuations is guaranteed when
\begin{eqnarray}
\left(\frac{\partial^2 \Phi}{\partial\rho_1^2}\right)
\left(\frac{\partial^2\Phi}{\partial  
\rho_2^2}\right)-\left(\frac{\partial^2\Phi}{\partial\rho_1\partial\rho_2}
\right)^2> 0.
\label{crite}
\end{eqnarray}
Fixing the areas of all species to one (in this case the demixing 
criterion will depend on the difference in particle shapes and not on 
their areas) the Eq. (\ref{crite}), in terms of the variables 
$(x,y)$, can be written as 
\begin{eqnarray}
H\equiv \frac{(1+y)^4}{y^2}\left[\frac{\partial}{\partial y}\left(
y^2\frac{\partial\varphi}{\partial y}\right)
\frac{\partial^2
\varphi}{\partial x^2}-\left(y\frac{\partial^2 \varphi}
{\partial x\partial y}\right)^2\right]>0,\nonumber\\
\end{eqnarray}
which, for the N phase and at the bifurcation point 
[with use of (\ref{uso})], becomes 
\begin{eqnarray}
&&H_{\rm{N}}^*=H_{\rm{I}}^*-\frac{\left(1+y^*\right)^4}{
2\left(y^*\right)^2B^*}\Bigg\{\left[\frac{\partial }{\partial y}
\left(y^2\frac{\partial\varphi_{\rm{I}}}{\partial y}\right)\right]^*
\left( A_x^*\right)^2\nonumber \\
&-& 2\left(\frac{\partial^2\varphi_{\rm{I}}}{\partial x\partial y}\right)^*
\left(y^*\right)^2 A^*_x A^*_y
+\left(\frac{\partial^2 \varphi_{\rm{I}}}{\partial x^2}\right)^*
\left(y^* A_y^*\right)^2\Bigg\}>0.\nonumber\\ 
\label{principal}
\end{eqnarray} 
It is easy to show that $H_{\rm{I}}(x,\eta)>0$ for HDR, freely rotating HR and 
HR in the Zwanzig approximation. 
This result confirms the general wisdom \cite{Talbot} mentioned above on
the stability of a two-dimensional 
mixture of hard bodies against I-I fluid demixing, in the framework of the SPT. 
However, fixing the particle orientations to be parallel we obtain
the condition for Eq. (\ref{crite}) to be violated as
\begin{eqnarray}
\eta\geq\eta^*\equiv \left[1+ x(1-x)
\left(\tau-\tau^{-1}\right)^2\right]^{-1/2},
\label{parallelNN}
\end{eqnarray} 
with
\begin{eqnarray}
\tau=\left\{\begin{array}{ll}\displaystyle\sqrt{\frac{\displaystyle\kappa_2-1+\frac{\pi}{4}}
{\displaystyle\kappa_1-1+\frac{\pi}{4}}},&\hbox
{for HDR},\\\displaystyle\sqrt{\frac{\kappa_2}{\kappa_1}},&\hbox{for HR},\end{array}\right.
\end{eqnarray} 
and, correspondingly, the position of the critical point in the pressure-composition
plane is
\begin{eqnarray}
x_c=\frac{1}{2},\hspace{0.4cm}\beta P_c v_1=\frac{ 3\left(\tau+\tau^{-1}\right)-2 }{
\left(\tau+\tau^{-1}-2\right)^2},
\end{eqnarray} 
where $\kappa_{\nu}=\left(L_{\nu}+\sigma_{\nu}\right)/\sigma_{\nu}$ for 
HDR while $\kappa_{\nu}=L_{\nu}/\sigma_{\nu}$ for HR (remember that these
results apply to mixtures such that all particle areas have been taken to be
unity). This result confirms
that there exists N-N demixing in the approximation that particles are
perfectly aligned. 

The condition $H_{\rm N}^*=0$
will be used in the next section to calculate 
the I-N demixing region in the plane 
$\lambda-x_1^*$, where $\lambda=\kappa_2/\kappa_1$ 
and $\kappa_{\nu}$ is the aspect ratio of species $\nu$, and
it is equivalent to the condition that the
Gibbs free energy per particle of the N phase have zero curvature 
(second derivative with respect to $x_1$) 
at bifurcation, thereby determining the I-N demixing tricritical point. 

\section{Results from the bifurcation analysis}
This section shows the results from the bifurcation analysis, whose formalism
was introduced in the preceding section. We have divided this section into
three sections. The first two are devoted to the study of the 
freely-rotating HDR and HR models, respectively, while in
the third section we discuss the implementation 
of the formalism to the restricted-orientation model (Zwanzig model). 
In this last section
results from the calculation of the exact phase diagrams of 
this model are shown. The main purpose is to check 
the results from the bifurcation analysis against exact calculations 
(note that there is no need to use any parametrization in the 
Zwanzig model).

\subsection{Hard discorectangles}
The solution of Eq. (\ref{uno}) for HDR allows us to write the 
packing fraction value at the bifurcation point as a function of the 
composition of the mixture,
\begin{eqnarray}
\eta^*=\left(1+\frac{2}{3\pi}\frac{\langle L^2\rangle}{\langle v\rangle}
\right)^{-1},
\label{inserting}
\end{eqnarray} 
where we have defined $\left<u^n\right>=\sum_{\nu}x_{\nu}^*u_{\nu}^n$ 
for a generic quantity $u$. The expression for $B^*$ which, as pointed 
out before, defines the 
relative positions of the I and N branches near the bifurcation point, is 
\begin{eqnarray}
B^*&=&\frac{x_1^*}{4^4z_1}[
z_2(3+z_1)r^2-2z_1z_2r+z_1(3+z_2)],\hspace*{0.1cm} 
\label{labast}
\end{eqnarray}
where $r=L_2/L_1$ and $z_{\nu}=x_{\nu}^*L_{\nu}^2/\langle L^2\rangle<1$. The 
quadratic polynomial $P(r)$ enclosed by the square brackets has a
discriminant $D=-48z_1z_2$, which obviously is always negative, while 
$P(0)>0$. Thus the coefficient $B^*$ is always positive and, as a 
consequence, the nematic branch always bifurcates from below with respect to the 
isotropic one. 

We take the areas of all particles to be equal to 1, which means that the
particle length and width, in units of $v_{\mu}^{1/2}$, are
\begin{eqnarray}
\sigma_{\mu}=\left(\kappa_{\mu}-1+\frac{\pi}{4}\right)^{-1/2},\hspace{0.4cm}
L_{\mu}=\left(\kappa_{\mu}-1\right)\sigma_{\mu}.
\end{eqnarray}
Inserting (\ref{inserting}) and (\ref{labast}) 
in (\ref{principal}), the 
final expression for $H^*_{\rm{N}}$ can be written in a 
particularly compact form,
\begin{eqnarray}
H_{\rm{N}}^*&=&\frac{\left(1+y^*\right)^4}{x_1^*x_2^* \left(y^*\right)^2}
\Bigg[4+\frac{3\pi}{\langle L^2\rangle}\nonumber \\
&-&\frac{8}{3}s_3^2
\left(1+s_2^2+\frac{3}{4}\frac{\langle p\rangle^2}{\langle L^2\rangle}
(s_2-s_1)^2\right)\Bigg], \label{final0}\\
s_1&=&\sqrt{\frac{\langle p^2\rangle}{\langle p\rangle^2}-1}, 
\quad s_2=\sqrt{\frac{\langle L^4\rangle}{\langle L^2\rangle^2}-1},\\
s_3&=& \sqrt{\frac{3\langle L^2\rangle^3}{4\langle L^4\rangle\langle 
L^2\rangle-\langle L^3\rangle^2}},
\end{eqnarray}
where $p=2L+\pi\sigma$ is the perimeter of the particle.

\begin{figure}
\mbox{\includegraphics*[width=3.5in, angle=0]{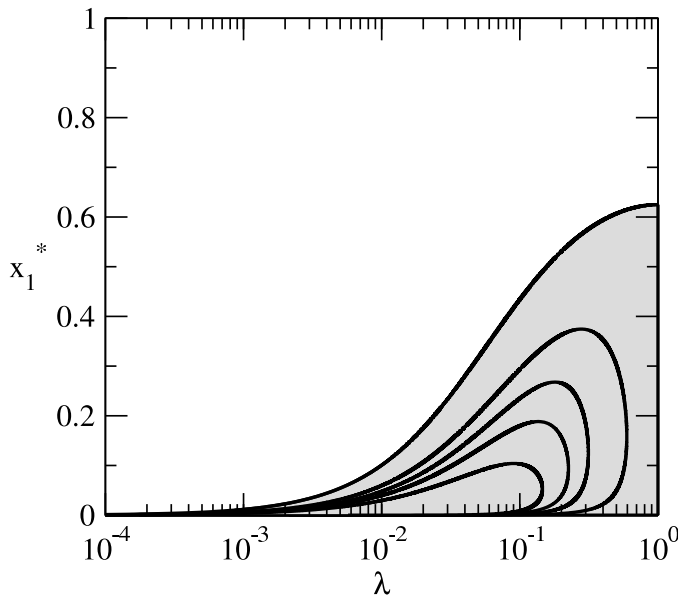}}
\caption{ Solutions of $H_{\rm{N}}^*=0$ in the $x_1^*$ (composition)-$\lambda$ plane,
where $\lambda=\kappa_2/\kappa_1$, in logarithmic scale, for HDR mixtures
with particle areas equal to unity.
The curves are shown for different values of $\kappa_2$. The shaded areas
represent the demixing regions predicted from $H_{\rm{N}}^*<0$. From outside
to inside the values of $\kappa_2$ are 1, 2, 3, 5, and $\infty$
(the Onsager limit).}
\label{fig1}
\end{figure}

The equality to zero of the expression enclosed 
by square brackets in (\ref{final0}), for the specific 
case $L_2=0$ (or $\kappa_2=1$, i.e. a binary mixture of hard disks and HDR), gives us 
the analytic solution $x_1^*=15\pi/4(6\pi+L_1^2)$,
shown in Fig. \ref{fig1} as the curve which encloses all the other 
curves in the $\lambda-x_1^*$ plane ($\lambda=\kappa_2/\kappa_1$).
Note that the above expression is not valid when $\lambda=1$ or $x_1^*=0$ 
(the one-component hard disk fluid). 
In Fig. \ref{fig1} the solutions of $H_{\rm{N}}^*=0$ for other mixtures, 
including the Onsager limit (i.e. a mixture of hard needles), are shown. It is 
remarkable that, even in this limit, the mixture can demix. To calculate 
these curves the aspect ratio $\kappa_2$ was fixed, while 
$\kappa_1$ was varied 
from 1 to $10^4$, and $H_{\rm{N}}^*=0$ was solved for $x_1^*$. The curve 
corresponding to the Onsager limit $\kappa_{\mu}\to \infty$ ($\mu=1,2$)   
was calculated by first taking the corresponding limit of the 
expression enclosed by square brackets in Eq. (\ref{final0}) (which 
depends only on $\lambda$) and solving again for $x_1^*$.
\begin{figure}
\mbox{\includegraphics*[width=3.5in, angle=0]{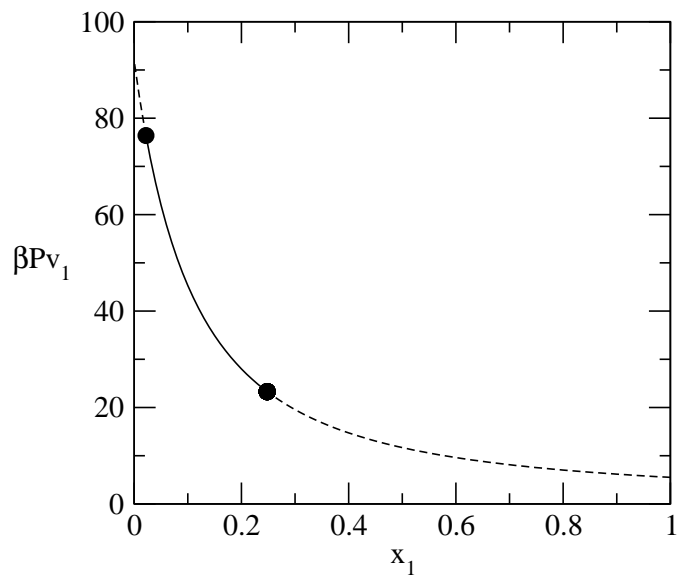}}
\caption{Reduced pressure $\beta Pv_1$ versus composition $x_1$ for a
binary mixture of HDR with $\kappa_1=5$ and $\kappa_2=2$. Filled circles
indicate the intersections of a vertical line located at the corresponding
value of $\lambda=2/5$ and the solution of $H_{\rm{N}}^*=0$ in Fig. \ref{fig1}.
The continuous line is the portion corresponding to the shaded region in
the latter figure.}
\label{fig2}
\end{figure}

We now proceed to explain the physics behind the behavior shown in 
Fig. \ref{fig1}
using the particular values $\lambda=2/5$, $\kappa_2=2$ as an example. A vertical line
located at this value of $\lambda$ intersects the curve corresponding to 
$\kappa_2=2$ at two points, giving two values of $x_1^*$ at which $H_{\rm{N}}^*=0$.
In Fig. \ref{fig2} we plot the reduced pressure as a function of $x_1$ for 
these values of $\lambda$ and $\kappa_2$. As can be seen from this figure the
pressure is a monotonically decreasing function of $x_1$, which means that
the above-mentioned intersection with a higher value of $x_1^*$ has a lower
pressure than the other one. This, in turn, implies that the first point
corresponds to a genuine tricritical point where, for the first time, the
Gibbs free energy loses its convexity with respect to the composition variable.
The second point also corresponds to a loss of convexity of the Gibbs free 
energy but, as will be discussed later, lies inside the two-phase region of
a demixing transition.

\subsection{Hard rectangles}

As shown in Refs. \cite{Martinez-Raton2} and \cite{Schaklen} the system 
of HR exhibits a transition to a phase with fourfold symmetry, 
the so-called tetratic phase (N$_t$). 
In this phase the orientational distribution 
function has a symmetry under rotation by $\pi/2$, $h(\phi)=h(\pi/2-\phi)$. 
The I-N$_t$ transition was always found to be of second order in the whole 
region of its stability ($1\leq \kappa \leq 2.21$).
Thus the question naturally arises as to the relative stability of this 
phase in the binary mixture. The above symmetry dictates  
that the odd Fourier amplitudes $\{h^{(\nu)}_{2j-1}\}$ should be equal to zero. Using 
this constraint and carrying out the same expansion for the free-energy 
difference around the bifurcation point, we arrive at the same Eq.
(\ref{primero}), with the coefficients $a_{\mu\nu}^{(k)}$ ($k=1,2$) obtained 
from Eq.(\ref{first}), but through the substitution $k\to 2k$ on the 
right-hand side of (\ref{first}), while for $k=3,4$ the coefficients
are given by the same expression (\ref{second}). 
The packing fraction at the I-N$_t$ bifurcation point, obtained as the 
solution of (\ref{uno}), is
\begin{eqnarray}
\eta^*_{\rm{N}_t}=
\left(1+\frac{2}{15\pi}\frac{\langle(L+\sigma)^2\rangle}{\langle v
\rangle}\right)^{-1}, 
\label{tetratic}
\end{eqnarray} 
while the value corresponding to the isotropic-uniaxial nematic (N$_u$) 
bifurcation point, calculated directly from (\ref{first}), (\ref{second}), 
and (\ref{uno}), is 
\begin{eqnarray}
\eta^*_{\rm{N}_u}=\left(1+\frac{2}{3\pi}\frac{\langle(L-\sigma)^2\rangle}
{\langle v\rangle}\right)^{-1}.
\label{uniax}
\end{eqnarray} 
The equality of (\ref{tetratic}) and (\ref{uniax}) defines a line 
in the plane $x_1^*$-$\kappa_1$ (fixing $\kappa_2$) where the 
I-N$_t$ transition 
preempts the I-N$_u$ transition. Taking the areas of both particles 
to be unity, as was done for HDR, and changing to variables 
$\theta_{\nu}=
\ln(\kappa_{\nu})/2$ ($\nu=1,2$), we obtain the solution of 
$\eta_{\rm{N}_t}=\eta_{\rm{N}_u}$ as 
\begin{eqnarray}
x_1^*=\frac{1-4\sinh^2\theta_2}{4\left(\sinh^2\theta_1-\sinh^2\theta_2
\right)},
\label{obtuvimos}
\end{eqnarray}  
with the constraint $\sinh^2\theta_{\mu}\leq 1/4\leq \sinh^2\theta_{\nu}$ 
(and $\mu\neq \nu$). 

This result should be taken with some care because the 
I-N$_\alpha$ ($\alpha=u,t$) transitions can be of first order and the 
relative position of binodals can change the scenario predicted above.
In order to elucidate the nature of these transitions, we have first
calculated the coefficient $B^*_{\rm{N}_t}$ for the tetratic 
phase which results in 
\begin{eqnarray}
B^*_{\rm{N}_t}=
\frac{x_1^*}{4^5z_1}\left[z_2(11+5z_1)r^2-10z_1z_2r+z_1(11+5z_2)
\right],\nonumber\\
\end{eqnarray}
where $r=\cosh\theta_2/\cosh \theta_1$ and 
$z_{\mu}=x_{\mu}^*\cosh^2\theta_{\mu}/
\langle \cosh^2\theta\rangle$. Again the second-order polynomial with 
respect to $r$ enclosed by the square brackets is always greater than zero. 
Thus $B^*>0$ and the N$_t$ energy branch bifurcates from below from 
the I branch. In a second step we have calculated $H_{\rm{N}_t}^*$ which 
gives
\begin{eqnarray}
H_{\rm{N}_t}^*&=&16\frac{\left(1+y^*\right)^4}{x_1^*x_2^*\left(y^*\right)^2}
\Bigg[1-\frac{2}{11}s_3^2\nonumber \\
&\times&\left(1+s_2^2+15\frac{(s_2-s_1)^2}{1+s_1^2}
\right)\Bigg], \\
s_k&=& \sqrt{\frac{\langle \cosh^{2k}\theta\rangle}
{\langle \cosh^k\theta\rangle^2}-1}, \quad k=1,2, \\
s_3&=&\sqrt{\frac{11\langle \cosh^2\theta\rangle^3}
{16\langle \cosh^4\theta\rangle\langle\cosh^2\theta\rangle-
5\langle \cosh^3\theta\rangle^2}}.
\end{eqnarray}
\begin{figure}
\mbox{\includegraphics*[width=3.3in, angle=0]{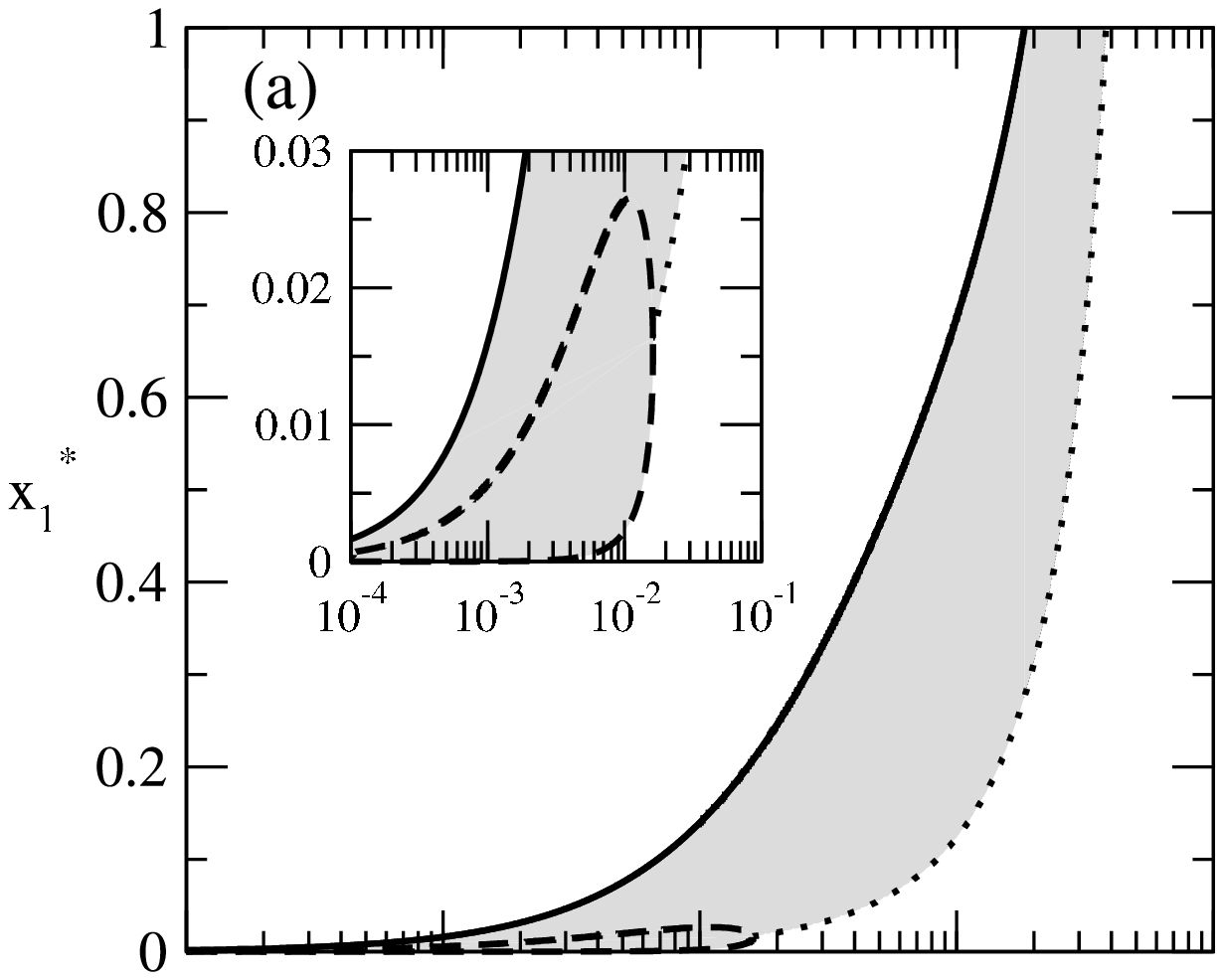}}
\mbox{\includegraphics*[width=3.3in, angle=0]{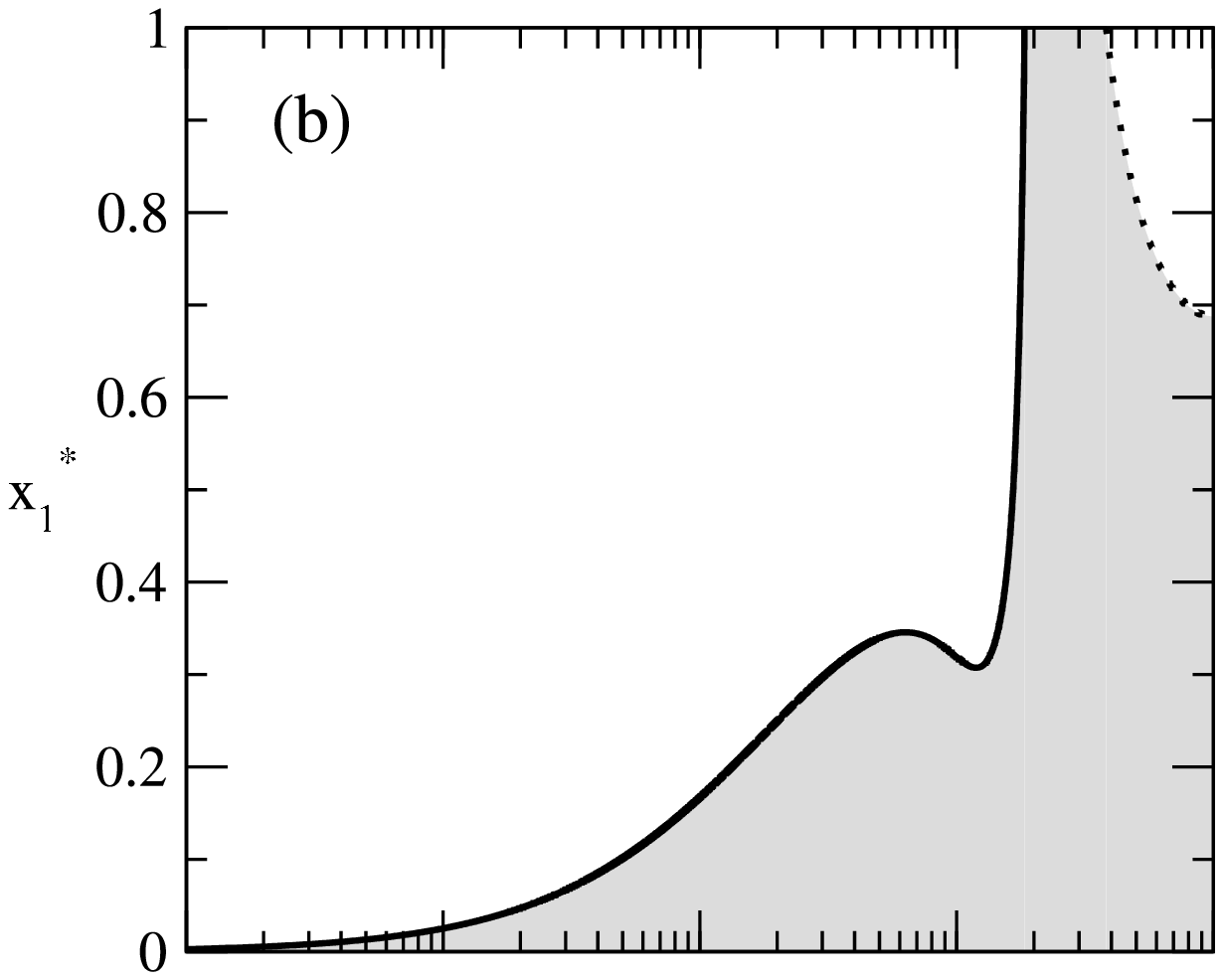}}
\hspace*{0.06cm}
\mbox{\includegraphics*[width=3.4in, angle=0]{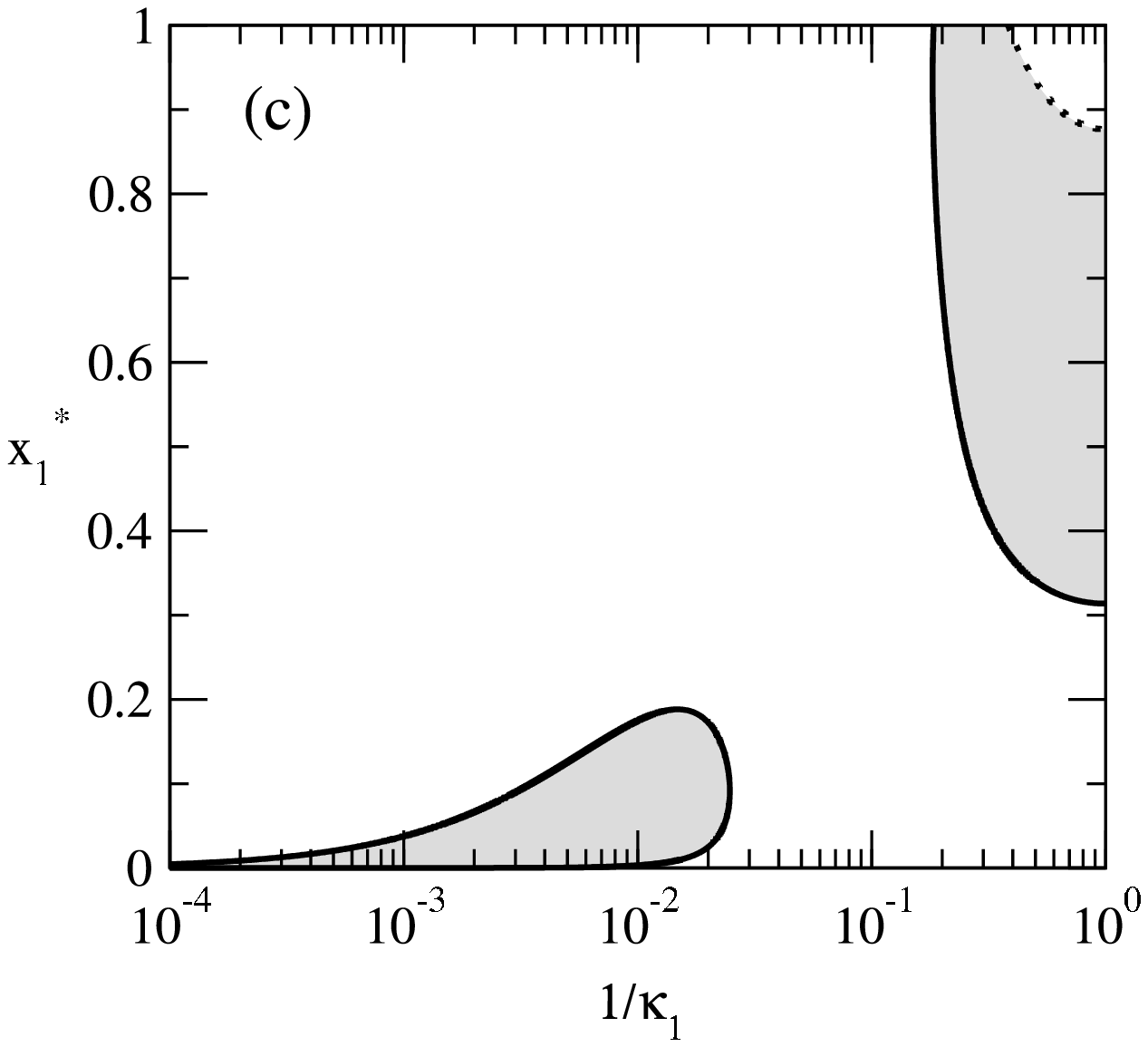}}
\caption{ Demixing regions predicted from $H_{\rm{N}_{\alpha}}^*=0$
($\alpha=u,t$) for HR in the $x_1^*-\kappa_1$ plane.
The solutions of the equations $H^*_{\rm{N}_u}=0$ (solid line),
$H^*_{\rm{N}_t}=0$ (dashed line), and the function (\ref{obtuvimos})
(dotted lines) are shown for
$\kappa_2=1$ (a), 5 (b), and 10 (c).}
\label{fig3}
\end{figure}

The region in the $x_1^*-\kappa_1$ plane where $H_{\rm{N}_t}^*<0$ defines
the demixing region, bounded by a dashed line in the inset 
of Fig. \ref{fig3}(a) for $\kappa_2=1$. As can be seen this region is 
closed as in the case of HDR for the I-N transition. The main reason for 
this behavior is that in the one-component limit the I-N$_t$ transition is 
of second order ($B^*_{\rm{N}_t}>0$ and 
$\left[\left(\varkappa_T\right)^{-1}_{\rm{N}_t}\right]^*>0$ 
\cite{kappa} for all 
$\kappa_1$) 
which in turn means that the solutions of 
$H_{\rm{N}_t}(x_1^*,\kappa_1)=0$ do not intersect 
the lines $x_1^*=\{0,1\}$. 

As the third and last step we have calculated the expressions for 
$B^*_{\rm{N}_u}$ and $H_{\rm{N}_u}^*$, which are too large to be shown here,  
except for their one-component limits $x_1^*\to 1$,
\begin{eqnarray}
\lim_{x_1^*\to 1} B^{*}_{\rm{N}_u}
&=&\frac{1}{64}
\frac{5-2\coth^2\theta_1}
{5-\coth^2\theta_1}, 
\label{prima}\\
\left[\left(\varkappa_T\right)^{-1}_{\rm{N}_u}v_1 \right]^*&=&
y^*(1+y^*)\frac{15\coth^2\theta_1-6\coth^4\theta_1-5}
{5-2\coth^2\theta_1},\nonumber\\
\label{segu}
\end{eqnarray}
where, rather than $H$,  the inverse isothermal compressibility 
$\varkappa_T^{-1}$ was chosen as the 
adequate thermodynamic variable to elucidate the nature of the I-N$_u$ 
transition 
of a one-component fluid \cite{kappa}.
An important difference to be mentioned is that, for HR, $B^*_{\rm{N}_u}$ in 
the one-component limit can 
be negative for some $\kappa_1$ [see Eq. (\ref{prima})], which 
indicates the presence of a tricritical point. In this 
limit the condition 
$\left[\left(\varkappa_T\right)^{-1}_{\rm{N}_u}\right]^*=0$ 
is more stringent to determine 
the exact location of this point which gives, equating the 
numerator of the right-hand side of Eq. (\ref{segu}) to zero, the solution 
\begin{eqnarray}
\kappa_1^*=\frac{\sqrt{1+\sqrt{7/15}}+2/\sqrt{5}}{
\sqrt{1+\sqrt{7/15}}-2/\sqrt{5}}\approx5.44 \nonumber 
\end{eqnarray}
\cite{Schaklen}.
On the other hand, under the assumption that both the I-N$_u$ and 
I-N$_t$ are of second order, the transition to the 
N$_t$ phase preempts the I-N$_u$ transition for $\kappa_1$ less than a 
value $\kappa_1^*$ obtained from Eq. (\ref{obtuvimos}) by 
taking the one-component 
limit $x_1^*=1$. The result is 
$\kappa_1^*=\left(3+\sqrt{5}\right)/2\approx 2.62$. 
However, for this value of $\kappa_1^*$, the I-N$_u$ transition is
of first order, so that the value of $\kappa_1^*$ that determines 
when the N$_t$ phase begins to be stable (for $\kappa_1<\kappa_1^*$)
should be found as the intersection of the I-N$_t$ spinodal with the I 
binodal of the I-N$_u$ coexistence. This value turns out to be 
\cite{Martinez-Raton2} $\kappa_1^*=2.21$. Thus, in the one-component limit, 
the range where the transition from the I phase to an orientational ordered 
phase is of first order is $2.21\alt \kappa_1\alt 5.44$.
For the mixture, the more stringent condition turns out to be  
$H_{\rm{N}_u}^*(x_1^*,\kappa_1)=0$ whose solutions are shown  
for $\kappa_2=1$, 5, and 10 (see Fig. \ref{fig3}). 
The main differences between these figures and that obtained for HDR can 
be summarized as follows: (i) 
Some of the  predicted demixing regions 
shown in Fig. \ref{fig3} are open due to the 
first order nature of the I-N$_u$ transition 
in the one-component limit, as was discussed above. In contrast the 
solution to $H_{\rm{N}_t}^*=0$ generates a closed loop due to the second order 
nature of the I-N$_t$ transition [see the inset of Fig. \ref{fig3}(a)].
Finally for $\kappa=10$ [Fig. \ref{fig3}(c)]  
$H_{\rm{N}_u}^*=0$ has two separate solutions, one of which is closed,  
the other one bounding an open region. In the Onsager limit we again obtain the 
closed loop shown in Fig. \ref{fig1}, as should be expected, since both models 
have the same asymptotic limit.   
(ii) Due to the presence of the N$_t$ phase in the mixture of HR,
one of the curves which 
bounds the demixing regions is given by Eq. (\ref{obtuvimos}) 
(the dotted lines in Fig. 
\ref{fig3}). (iii) The demixing regions for HR mixture are in general 
wider as compared to those of HDR.

Let us discuss the behavior of the pressure, using Fig. \ref{fig3}(a) 
and, as an example, the case 
 $\kappa_2=1$ and $\kappa_1=10^2$. 
This behavior is similar to that shown 
in Fig. \ref{fig2} for the HDR model, namely the pressure is a monotonically
decreasing function of composition. 
Also there are two points obtained from the intersection of the vertical line 
at $\kappa_1^{-1}=10^{-2}$ and the boundaries of the shaded region 
(the upper one on the solid line and the other on the lower branch of the 
dashed line). These in principle would correspond to  tricritical points 
as in the case of HDR. The point corresponding to a lower pressure 
(higher composition) is a genuine tricritical point while the other 
one might be inside  I-N$_u$ or I-N$_t$ demixing regions.

\subsection{The Zwanzig model}

This simple model, as applied to  HR, 
allows the calculation of the phase diagram without any 
parametrization because the orientational distribution function can be 
taken as $h_{\mu}(\phi)=\left[\left(1+q_{\mu}\right)/2\right]\delta(\phi)+
\left[\left(1-q_{\mu}\right)/2\right]\delta(\phi-\pi/2)$ 
[$\delta(x)$ is the Dirac delta function], corresponding to a binary 
mixture of two species with perpendicular orientations. Thus, the excess part 
of the free 
energy is a second degree polynomial of the order 
parameters $q_{\nu}$ (with $-1\leq q_{\nu}\leq 1$), and 
the minimization of the total free energy requires to solve two transcendental 
equations to find their equilibrium values. Since no parametrization 
is necessary, the location of all tricritical points, as obtained from
the minimization, are exactly the same as those obtained from the
bifurcation analysis. This in fact is the reason why we have chosen to
use this model.

\begin{figure}
\mbox{\includegraphics*[width=3.5in, angle=0]{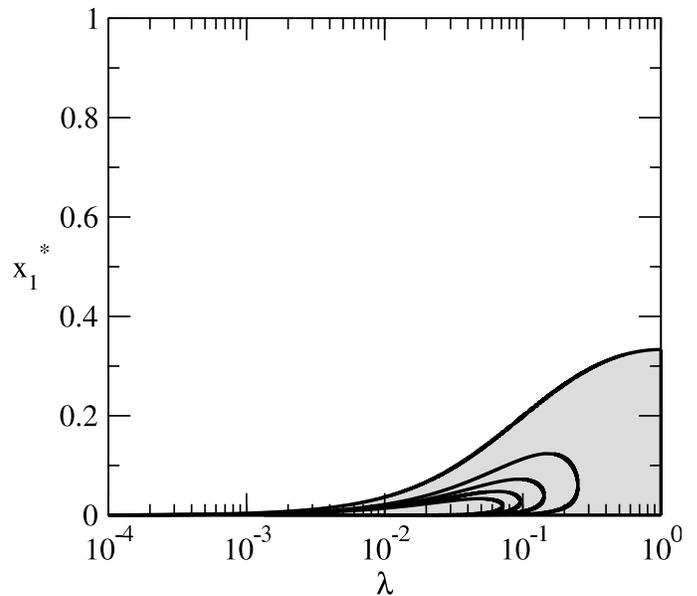}}
\caption{Demixing regions (shaded areas) of the Zwanzig model predicted
from $H_{\rm{N}}^*=0$.
Different curves are
the solutions of (\ref{zwanzig}) for
$\kappa_2=1$, 2, 3, 5, and $\infty$ (the Onsager limit) from outside
to inside.}
\label{fig4}
\end{figure}

Once we calculate the phase diagram we can compare the 
results to those obtained using the bifurcation analysis. 
To implement the latter we need the following expressions: 
\begin{eqnarray}
a^{(1)}_{\mu\nu}&=&\frac{x_{\mu}}{2}\left(\delta_{\mu\nu}
-2yx_{\nu}\sinh\theta_{\mu}\sinh\theta_{\nu}\right), \\
a^{(2)}_{\mu\nu}&=&a_{\mu\mu}^{(3)}=0,\quad a_{\mu\mu}^{(4)}=
\frac{x_{\mu}}{12},
\end{eqnarray} 
which allows, using the same procedure described above, to find the 
packing fraction at the bifurcation point,
\begin{eqnarray}
\eta^*=\left(1+2\langle \sinh^2\theta\rangle\right)^{-1},
\end{eqnarray}
and the coefficient $B^*$,
\begin{eqnarray}
B^*=\frac{1}{12}\frac{\langle \sinh^4\theta\rangle}{\sinh^4\theta_1}>0,
\end{eqnarray}
where the constraint of all particle areas being equal to 1 
was imposed and the 
same change of variables [$\theta_{\mu}=\ln(\kappa_{\mu})/2$] was used.
Finally, to find the demixing behavior of this model,
we need to make the expression 
\begin{eqnarray}
H_{\rm{N}}^*&=&\frac{\left(1+y^*\right)^4}{x_1^*x_2^*\left(y^*\right)^2}
\Bigg[1+a
-\frac{3}{2}\left(1+s_3^2\right)^{-1}\nonumber \\
&\times& \left(1+(as_2)^2+a
\frac{(as_2-s_1)^2}{1+s_1^2}\right)\Bigg],
\label{zwanzig} \\
s_k&=&\sqrt{\frac{\langle \cosh^{2k}\theta\rangle}{\langle\cosh^k
\theta\rangle^2}-1}, \quad k=1,2\\
s_3&=&\sqrt{\frac{\langle\sinh^4\theta\rangle}{\langle\sinh^2\theta\rangle^2}
-1},\quad a=\frac{\langle\cosh^2\theta\rangle}{\langle\sinh^2\theta\rangle}
\end{eqnarray}
to vanish. The solutions of this equation  for different values of $\kappa_2$
are shown in Fig. \ref{fig4}. 
As can be seen the general topology is similar to that found for HDR. 
Comparing both figures (\ref{fig1} and \ref{fig4}) we can draw 
as a conclusion  
that the discretization of orientations unfavors the I-N demixing.

\begin{figure}
\mbox{\includegraphics*[width=3.5in, angle=0]{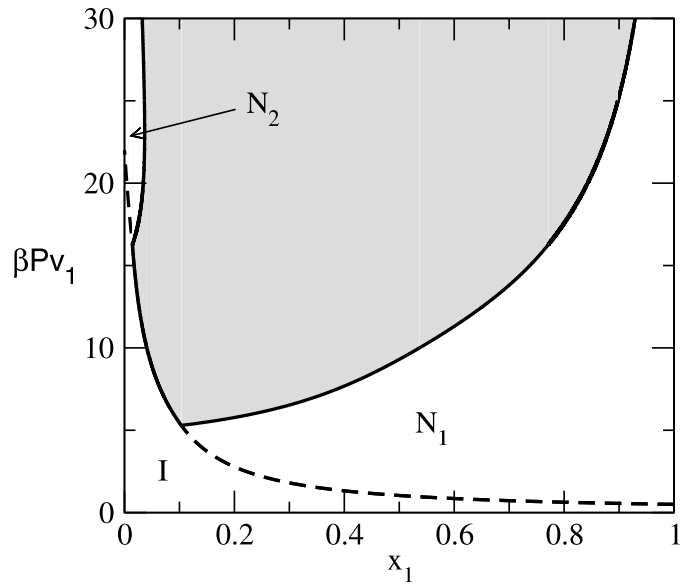}}
\caption{Phase diagram of the Zwanzig mixture with $\kappa_1=9$ and
$\kappa_2=2$.}
\label{fig5}
\end{figure}
   
The phase diagram of a Zwanzig binary mixture of species with aspect 
ratios $\kappa_{\mu}$ equal to 9 ($\mu=1$) and 2 ($\mu=2$) 
is shown in Fig. \ref{fig5} in the pressure-composition plane.
The tricritical point (where the 
I-N$_1$ demixing transition changes from first to second order as the 
pressure is reduced)
at $x_1^*=0.1034$ and $\beta P^*v_1=5.2928$ 
coincides exactly with 
the point predicted from the bifurcation analysis (see in Fig. \ref{fig4} 
the upper intersection of the curve for $\kappa_2=2$ with a 
vertical line located 
at $\lambda=2/9\approx 0.222$). 

An important feature of this 
phase diagram is that the binodals of the I-N$_1$ or N$_1$-N$_2$ 
transitions tend to the one-component asymptotes $x_1=0$ and $x_1=1$ as the 
pressure is increased. This trend, characteristic of demixing scenarios 
(such as the usual I-I fluid demixing), should be compared with other possible
topologies of phase diagrams (for example, those of three-dimensional 
mixtures of particles of similar lengths),
where the first-order I-N transition has two binodals 
that meet at the one-component limits of the phase diagram {\it at finite
pressure}. This criterion is the one we have used to consider
the transitions found here as I-N demixing transitions instead of the 
standard I-N orientational transition. 

\begin{figure}
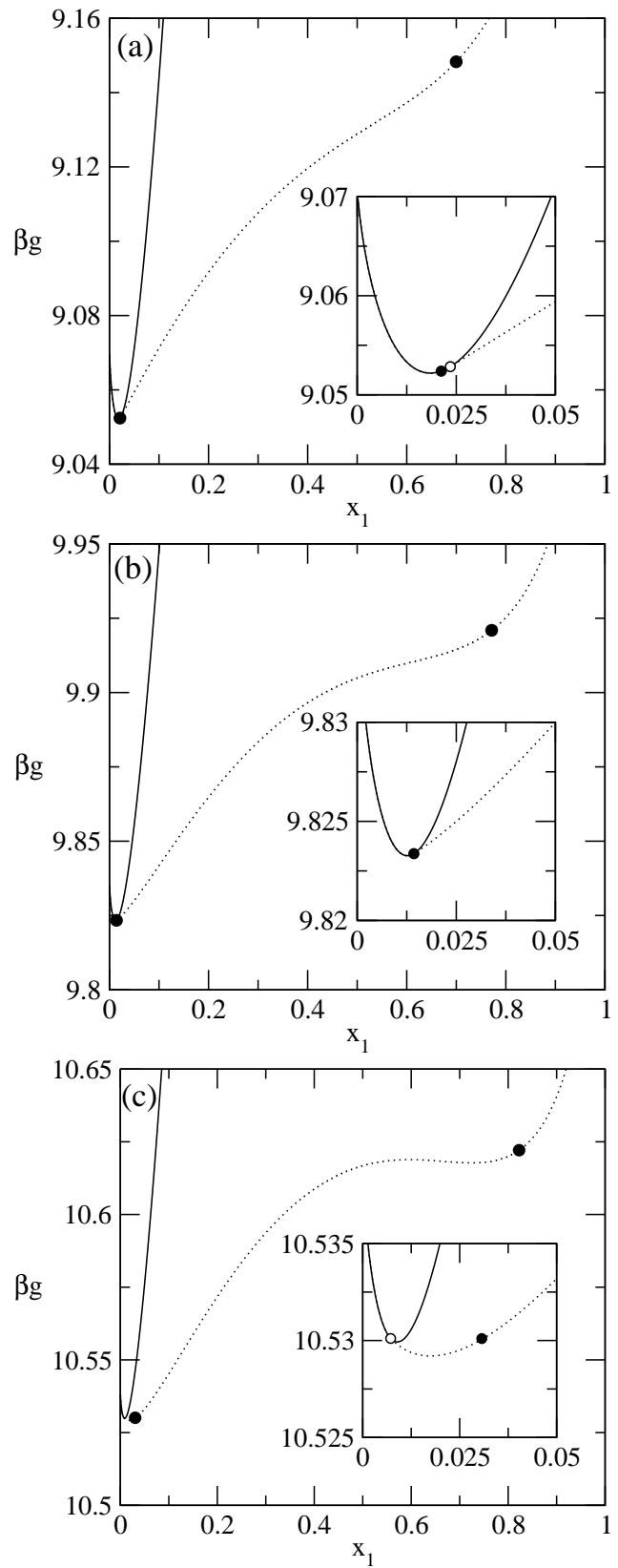

\mbox{\includegraphics*[width=3.3in, angle=0]{fig6a.eps}}
\mbox{\includegraphics*[width=3.3in, angle=0]{fig6b.eps}}
\mbox{\includegraphics*[width=3.3in, angle=0]{fig6c.eps}}
\caption{Gibbs free energy per particle in reduced units
$\beta g$ as a function of $x_1$
for the Zwanzig mixture with species having
$\kappa_1=9$ and $\kappa_2=2$. The values of the pressure are
$\beta Pv_1=13.8014$ (a), $\beta Pv_1=16.2775$ (b), and
$\beta Pv_1=18.7536$ (c). The solid and dotted lines correspond to the I
and N branches, respectively.
The filled circles represent the coexistence points. The
insets are enlargements of the neighborhood of
the bifurcation points, which are shown by open
circles. Panel (a) shows the loss of convexity of $\beta g$ in the N brach
at the bifurcation point.}
\label{fig6}
\end{figure}

As can be seen from Fig. \ref{fig5}, at higher pressure from the tricritical 
point there is a triple intersection between two first order transition 
lines (the I binodal of the I-N$_1$ transition and the N$_2$ binodal of 
the N$_1$-N$_2$ transition) and the second order I-N$_2$ transition line. 
This point is called in the literature a \emph{critical endpoint}.
An important remark to make is that 
this critical endpoint located at 
$x_1^*=0.0144$ and $\beta P^*v_1=16.2775$ 
does not exactly coincide with 
the predicted result from the bifurcation analysis.  
If we look at the lower intersection between 
the curve corresponding to $\kappa_2=2$ and the vertical line 
located at $\lambda\approx 0.222$ in Fig. \ref{fig4}, 
we find the value $x_1^*$ which corresponds to 
$\beta Pv_1=13.8014$, which  
is lower than the above mentioned critical endpoint pressure.  
The reason for this disagreement can be elucidated if we plot the Gibbs 
free energy per particle of the mixture as a 
function of $x_1$ for three different values of the pressure,
$\beta Pv_1=13.8014$, which is the pressure corresponding to 
the upper tricritical point as obtained from the bifurcation
analysis [the open circle in the inset of Fig. \ref{fig6}(a)
corresponds to the bifurcation point];
$\beta Pv_1=16.2775$, which corresponds to the triple intersection
[Fig. \ref{fig6}(b)] where the bifurcation and the coexistence
points coalesce; and $\beta Pv_1=18.7536$, where the two coexistence
points already lie on the nematic branch. 

The main conclusion we can draw from this 
is that the bifurcation 
analysis predicts exactly the tricritical point 
(which corresponds to the lower 
pressure or higher $x_1^*$ in Fig. \ref{fig4}) and 
gives an approximate value for the critical endpoint. 

In the light of these results we can interpret Fig. \ref{fig3}(a) and try 
to predict, as an example, the demixing behavior of a mixture of 
freely rotating HR with
$\kappa_2=1$ and $\kappa_1= 10^2$. If we increase the pressure 
$\beta P v_1(x_1^*)$ following the I-N$_u$ spinodal curve 
we find that at a composition defined as the
intersection with the solid line (the tricritical point) 
the mixture begins to demix into I and 
N$_u$ phases. The demixing gap becomes wider with pressure, and at 
high pressures the lower intersection with the dashed line defines the 
approximate location of a critical endpoint where the first order 
I-N$_u$ and N$_u$-N$_t$ transitions coalesce with a second order 
I-N$_t$ transition. Increasing further the pressure we should have a 
N$_u$-N$_t$ coexistence and, ultimately, a coexistence between two uniaxial 
nematic phases. This latter coexistence could be predicted if we take into 
account the phase diagram for 
the one-component system, where the N$_t$ phase is sandwiched between the I 
and N$_u$ phases \cite{Martinez-Raton2}.   

\section{Phase diagrams of HDR mixtures}

In this section we want to explicity discuss the demixing scenarios 
that may occur
in binary mixtures of hard particles. We will focus our attention on 
HDR mixtures. For this purpose we have parameterized the 
orientational distribution functions of each species as 
\begin{eqnarray}
h_{\mu}(\phi)=\frac{1}{\pi I_0(\alpha_{\mu})}\exp{\left(\alpha_{\mu} 
\cos 2\phi\right)},
\end{eqnarray}
where $I_0(x)$ is the zeroth-order modified Bessel function. This 
parametrization fulfills the normalization constraint $\int_0^{\pi}
d\phi h_{\mu}(\phi)=1$. The free energy per volume (see Sec. 
II for its expression) was minimized with respect to $\alpha_{\mu}$ 
($\mu=1,2$) and the double-tangent construction on the thermodynamic 
potential $\beta g(x_1)$ (the Gibbs energy per particle in reduced 
units as a function 
of the composition of the mixture for a fixed pressure) was used 
to calculate the coexistence values of the composition of the mixture and 
the packing fraction.

\begin{figure}
\mbox{\includegraphics*[width=3.3in, angle=0]{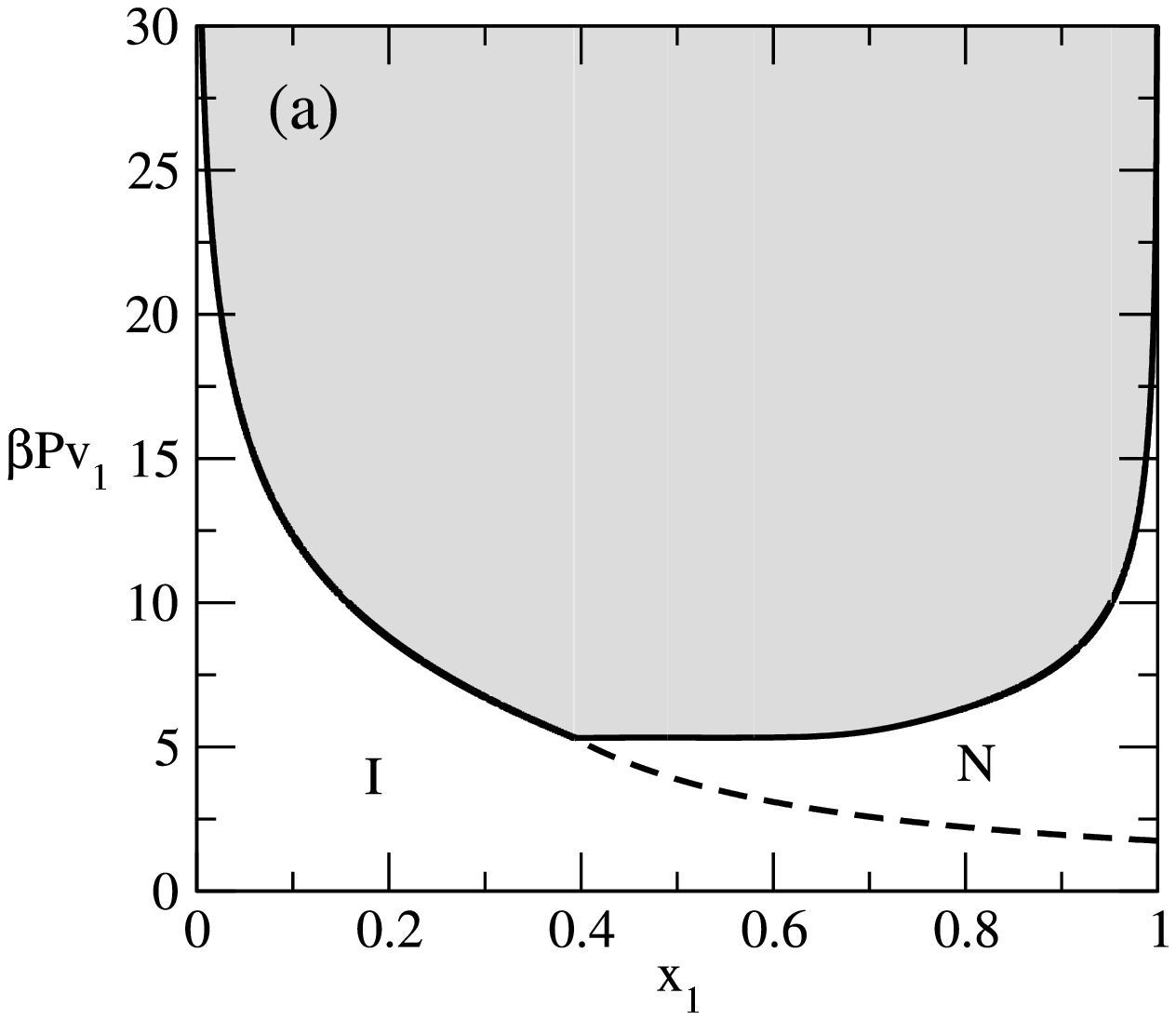}}
\mbox{\includegraphics*[width=3.5in, angle=0]{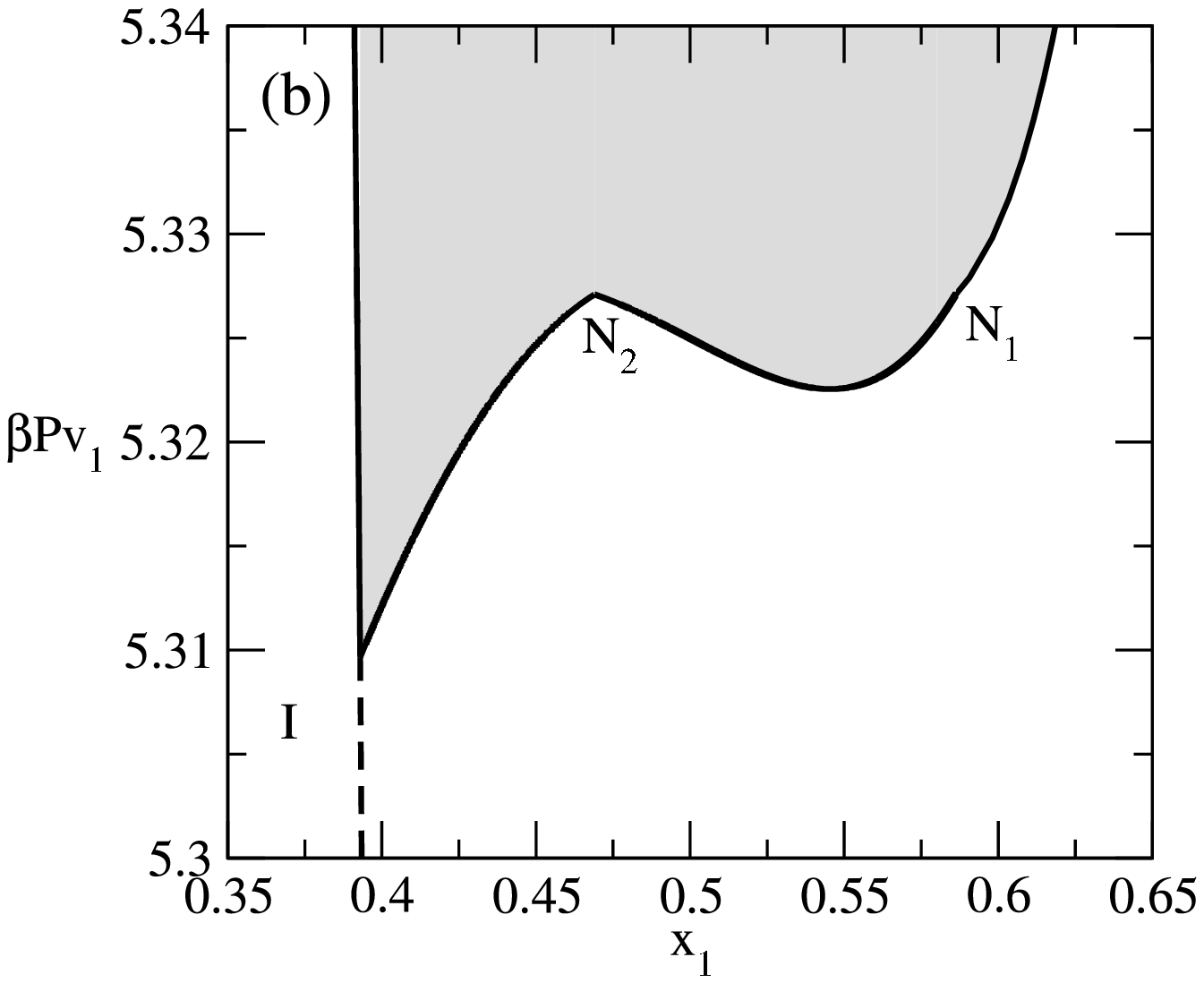}}
\caption{(a) Phase diagram of the mixture of hard disks $\kappa_2=1$ and
HDR with $\kappa_1=10$. (b) A zoom taken around the tricritical point.}
\label{fig7}
\end{figure}
\begin{figure}
\mbox{\includegraphics*[width=3.5in, angle=0]{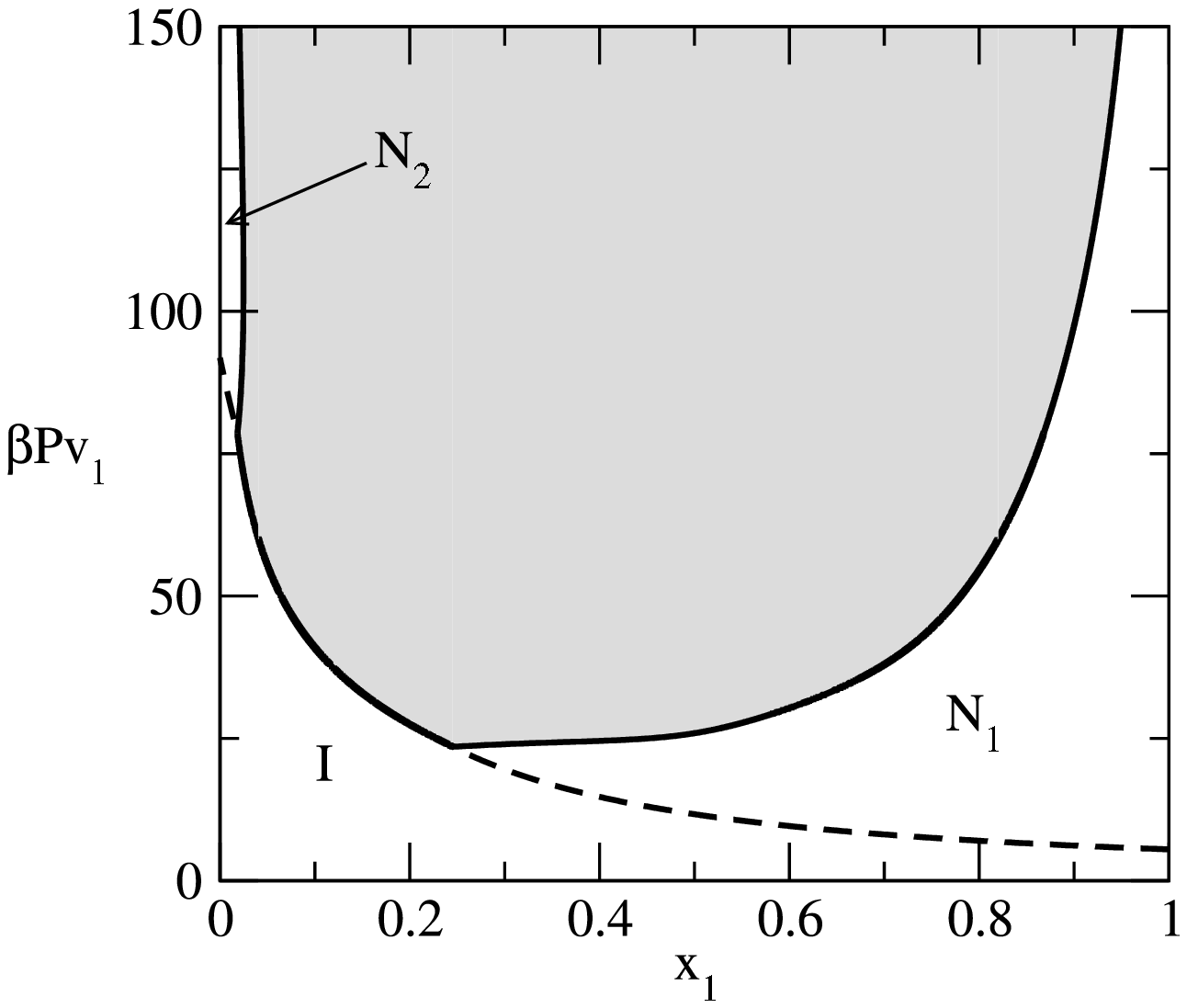}}
\caption{Phase diagram for a HDR mixture with $\kappa_1=5$ and $\kappa_2=2$.}
\label{fig8}
\end{figure}
\begin{figure}
\mbox{\includegraphics*[width=3.5in, angle=0]{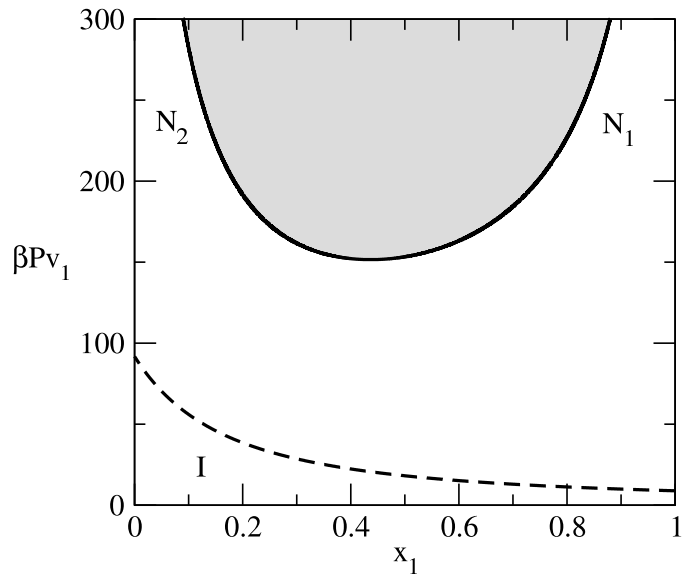}}
\caption{Phase diagram for a HDR mixture with $\kappa_1=4$ and $\kappa_2=2$.}
\label{fig9}
\end{figure}

The results for a mixture of hard disks ($\kappa_2=1$) and HDR 
with $\kappa_1=10$ are shown in Figs. \ref{fig7}(a) and \ref{fig7}(b). 
For low pressures there is a continuous I-N transition at a pressure which 
agrees with that calculated from the spinodal packing fraction curve 
(\ref{inserting}). At some composition ($x_1^*=0.393$) we find  
a tricritical point from which the mixture begins 
to demix into the I phase (rich in discs) and 
the N phase (rich in rods). The
location of the tricritical point should be compared with that resulting 
from the bifurcation analysis ($x_1^*=0.434$) which gives the exact result. The 
difference is due to the parametrization used. As we have already 
pointed out the value of $x_1^*$ depends on the coefficient $B^*$ which in turn 
depends on the $a_{\mu\nu}^{(k)}$ ($k=1,\dots,4$). These are 
the expansion coefficients of the free-energy 
difference around the bifurcation point up to fourth order.
Although the used parametrization 
captures the right second order terms, further terms are only approximate 
due to the restriction on the minimization variables to be only 
one per species.

An interesting result is shown in Fig. \ref{fig7}(b), a zoom of Fig. 
\ref{fig7}(a) around the tricritical point. As we can see this part of 
the phase diagram exhibits a N$_1$-N$_2$ coexistence which ends 
in a critical point, along with 
the presence of a I-N$_1$-N$_2$ triple point. As these features are
similar to those observed in the Zwanzig mixture (not 
shown here), where no parametrization was used, 
we are confident that this scenario is qualitatively correct.    

Two other phase diagrams are shown in Figs. \ref{fig8} and \ref{fig9} for 
different mixtures. 
The topology of the first one is similar to that found in the Zwanzig 
mixture, where the I-N$_1$ demixing is followed by N$_1$-N$_2$ coexistence 
and the phase diagram includes one tricritical and one critical endpoint. 

The third phase diagram (see Fig. \ref{fig9}) includes an I-N transition which 
is always of second order and, at high pressures, a N$_1$-N$_2$ demixing 
ending in a critical point. This kind of phase diagram is typical of 
mixtures of particles with similar aspect ratios. 
Since the pressures at which the demixing transition occurs are rather high,
we expect this transition to be metastable with respect to phases with
partial or full spatial order. Comparing the phase diagram of Fig. \ref{fig9}
with the  
exact bifurcation analysis results (see  
Fig. \ref{fig1}), 
we can conclude that, even though I-N demixing is already allowed  
from $\lambda \leq 0.595$, the parametrization used changes the demixing 
behavior of the mixture. An exact free-energy  
minimization for the same value of $\kappa_2$ and for $\lambda >0.595$ 
should qualitatively give a phase diagram similar to that 
shown in Fig. \ref{fig9}.

\section{Conclusions}

While I-I demixing is forbidden in two-dimensional 
hard-body additive mixtures of anisotropic 
particles according to SPT \cite{Talbot}, 
we have shown in the present work that 
the inclusion of ordered phases with orientational symmetry breaking
changes completely the demixing scenario. The phase diagrams characteristic 
of these mixtures can exhibit I-N and N-N demixing. We have used a 
bifurcation analysis to demonstrate rigorously that I-N demixing occurs.
On the other hand, explicit calculations of the phase diagrams of HDR,
using an accurate parametrization, have been performed in order
to show the occurrence of N-N demixing. The simple structure 
of SPT allowed us to obtain analytically the  
stability criterion for the mixture as a function of packing fraction, 
composition and shape of the constituent particles. Thus, using this 
procedure, we can predict the I-N demixing scenarios for mixtures of
HDR and HR.

To show the relative stability of these demixed phases with respect to 
nonuniform phases (e.g., solid phases), we would 
need to carry out a full minimization, with respect to 
the density profile $\rho_{\mu}({\bf r},\phi)$ 
(which also depends on spatial variables), of a density functional constructed 
in such a way that it recovers the SPT in the uniform limit. This work 
is a task in progress. 

Finally, we expect that all the demixing scenarios
predicted here be confirmed by computer simulations of 
hard-body mixtures in two dimensions. These simulations are still lacking.

\section*{ACKNOWLEDGMENTS}
One of the authors (Y. M.-R.)
was supported by a Ram\'on y Cajal research contract from Ministerio
de Educaci\'on y Ciencia (Spain). This work is part of research projects
Nos. BFM2003-0180, BFM2001-0224-C02-01,
BFM2001-0224-C02-02, and BFM2001-1679-C03-02 of the Ministerio de
Educaci\'on y Ciencia (Spain).


\begin{references}
\bibitem{kappa}Note that in the one-component limit the 
condition $H^*=0$ should be substituted by 
 $\left(\varkappa_T^{-1}\right)^*=0$ 
[where $\varkappa_T^{-1}=\rho
\left(\partial \beta P/\partial \rho\right)_T$ is    
the inverse isothermal compressibility].
\end{references}
\end{document}